\documentclass[review]{elsarticle}
\usepackage{comment}
\usepackage{float}
\usepackage{subfig}
\usepackage[top=3cm,left=2.8cm,right=3.3cm,bottom=3cm]{geometry}
\usepackage{lineno,hyperref}
\modulolinenumbers[5]
\usepackage{color}
\usepackage{multirow}
\usepackage{amsmath}
\usepackage{soul}
\usepackage{amssymb}

\usepackage[final]{changes}
\definechangesauthor[name={Luis}, color=orange]{imm}
\setremarkmarkup{(#2)}

\journal{Journal of Acta Materialia}

\begin{document}

\begin{frontmatter}

\title{Computationally efficient phase-field studies combining simulation sampling and statistical analysis}

\author{Christian Schwarze$^a$, Reza Darvishi Kamachali$^{a,b,}$\footnote{Corresponding author. Email: kamachali@mpie.de, reza.darvishi@rub.de}, Markus K{\"u}hbach$^{b,c}$, Christian Mie{\ss}en$^c$, Marvin Tegeler$^a$, Luis Barrales-Mora$^{c,d}$, Ingo Steinbach$^a$, G{\"u}nter Gottstein$^c$}
\address{$^a$Interdisciplinary Centre for Advanced Materials Simulation (ICAMS), Ruhr-University Bochum, Germany\\$^b$Max-Planck-Institut f\"ur Eisenforschung, Max-Planck-Stra{\ss}e 1, 40237 D\"usseldorf, Germany\\$^c$Institute of Physical Metallurgy and Metal Physics, RWTH Aachen University, 52074 Aachen, Germany\\$^d$George W. Woodruff School of Mechanical Engineering, Georgia Institute of Technology, 2 Rue Marconi, Metz, France}

\begin{abstract}
\label{Abstract}
The trade-off between accuracy and computational cost as a function of the size and number of simulation boxes was studied for large-scale phase-field simulations. For this purpose, a reference simulation box was incrementally partitioned. We have considered diffusion-controlled precipitation of $\delta '$ in a model Al-Li system from the growth stage until early ripening. The results of the simulations show that decomposition of simulation box can be a valuable computational technique to accelerate simulations without substantial loss of accuracy. In the current case study, the precipitate density was found to be the key controlling parameter. For a pre-set accuracy, it turned out that large-scale simulations of the reference domain can be replaced by a combination of smaller simulations. This shortens the required simulation time and improves the memory usage of the simulation considerably, and thus substantially increases the efficiency of massive parallel computation for phase-field applications.
\end{abstract}


\end{frontmatter}

\newpage
\linenumbers

\section{Introduction}
\label{Sec_Intro}

The structural and many functional properties of materials are determined by their microstructure. Modern alloy design encompasses a detailed understanding of the microstructure and the processes of microstructure modification to tailor materials to specific applications. In these complex tasks, modelling and simulation are widely utilized to predict microstructure and property evolution of materials during their synthesis, processing and even operation. Many computer models utilized in materials science describe the microstructure by using a spatially resolved representative volume element (RVE). An RVE is the smallest statistical representation of a microstructure \cite{Kanit2003, Drugan1996} that samples all it's relevant features. Determining the size of the RVE is not trivial \cite{Gusev1997, Ostoja-Starzewski2006,Ostoja-Starzewski2002,Shan2002} because in most materials the microstructural features are heterogeneously distributed and occur at different length scales. Hence, establishing the optimal size of the RVE is a difficult task. In fact, it has been pointed out by many investigations \cite{Kremeyer1998,Marek2013,Mason2015,Torquato2013} that the more heterogeneous a microstructure is the finer it has to be resolved. An example is the problem of grain growth where despite the possibility for studying large-scale simulations \cite{Darvishi2012,Darvishi2015_2,Miyoshi2017} it was demonstrated that the size of the RVEs utilized in several grain growth simulations may not be enough to observe self-similar behaviour depending on the initial grain size distribution \cite{Miessen2017}. Evidently, the computational cost of representing a microstructure as a continuous and contiguous space is immense. This has rendered most of the algorithms memory intensive and even memory bound. To remedy this situation, it was recently proposed \cite{Kuehbach2016} to take advantage of supercomputers and make use of numerous solitary units to represent a microstructure. A solitary unit is a small RVE that itself is not statistically representative but once numerous units are considered they approach excellently reality. This idea was tested for the simulation of recrystallization with excellent results \cite{Kuehbach2016}. \\

The question posed in the present contribution is whether the same concept can be applied to more complex physical models e.g. phase transformations, where diffusion can impose long-range effects and a sub-division of a large RVE into solitary units may not be possible owing to these effects. This problem can only be studied by means of the phase-field model since diffusion is an important issue to consider \cite{STEINBACH1996135, PhysRevE.71.041609, MOELANS2008268, EGGLESTON200191}. In fact, the multi-phase-field approach \cite{steinbach2006,Steinbach2009,Chen2002,Provatas2011,Vondrous2013} has proven its capability for studying many phenomena in materials science such as grain growth \cite{Suwa2007,Darvishi2010,Darvishi2012,Darvishi2015_2}, recrystallization and texture evolution \cite{Suwa2008,Takaki2009,Darvishi2015_1} and particle pinning \cite{Schwarze2016}. Furthermore, multiple features of elasticity, diffusion, fluid flow, effect of external fields, and different interface phenomena can be conveniently integrated into the phase-field framework \cite{Jeong2001,Wang2002,Zhang2007,Shibuta2007} as well as mutual coupling between them \cite{Kamachali2017}. \\

In the present contribution, we performed phase-field simulations of diffusion-controlled precipitation and ripening combined with statistical sampling to study the trade-off between computational costs and accuracy when a reference system is incrementally partitioned. The results of the simulations are investigated in terms of precipitate size and spacing over the course of evolution. A systemic sampling and statistical averaging was applied to investigate the efficiency of these simulations in terms of computational costs as well as accuracy of the results with respect to a reference simulation. 

\section{Model description and simulation procedure}
\label{Sec_Model}

\subsection{Multi-phase-field model}
\label{Sec_PF}
The multiphase-field method \cite{Steinbach2009} is based on the description of the total free energy by integrating the interface free energy density $f^{IN}$ and the chemical free energy density $f^{CH}$ over a domain $\Omega$ following the sum constraint ($\sum_{\alpha=1}^N\phi_\alpha=1$) for all existing phase variables $\phi$:
\begin{equation}
\label{eqF}
 F=\int_\Omega\left(f^{IN}+f^{CH}\right)dV.
\end{equation}

The interface energy density is given as
\begin{equation}
\label{eqFin}
 f^{IN}=\sum_{\alpha=1}^N\sum_{\beta\neq\alpha}^N\frac{4\sigma_{\alpha\beta}}{\eta} \{-\frac{\eta^2}{\pi^2}\nabla\phi_\alpha\cdot\nabla_\beta+\phi_\alpha\phi_\beta\}
\end{equation}
in which $\sigma_{\alpha\beta}$ is the interface energy between phases $\alpha$ and $\beta$ and, $\eta$ is the interface width. The chemical free energy density is

\begin{equation}
\label{eqFch}
 f^{CH}=\sum_{\alpha=1}^N\phi_\alpha f_\alpha\left({c}_\alpha\right)+\mu \left[{c}-\sum_{\alpha=1}^N\left(\phi_\alpha{c}_\alpha\right)\right]
\end{equation}
where $f_\alpha(c_\alpha)$ is chemical free energy of phase $\alpha$, $\mu$ the chemical potential, $c_\alpha$ the phase concentration of phase $\alpha$ and $c$ is the total concentration fulfilling $c=\sum_{\alpha=1}^N\phi_\alpha c_\alpha$.

The evolution of the phase-field using Equations \ref{eqF}-\ref{eqFch} follows
\begin{eqnarray}
 \dot{\phi}_\alpha = & -\sum_{\beta=1}^N\frac{\mu_{\alpha\beta}}{N}\left(\frac{\delta}{\delta\phi_\alpha}-\frac{\delta}{\delta\phi_\beta}\right)F \nonumber \\
 = & \sum_{\beta=1}^N\frac{\mu_{\alpha\beta}}{N}\left[\sum_{\gamma=1\neq\beta}^N[\sigma_{\beta\gamma}-\sigma_{\alpha\gamma}]\left[\nabla^2\phi_\gamma+\frac{\pi^2}{\eta^2}\phi_\gamma\right]+\frac{\pi^2}{8\eta}\Delta g_{\alpha\beta}\right]
 \label{Eq.PFF}
\end{eqnarray}
with $\mu_{\alpha\beta}$ as the interface mobility and $\Delta g_{\alpha\beta}$ only as the chemical driving force \cite{Tiaden1998}, which is proportional to the undercooling towards the line which separates the fcc-aluminium from the two-phase region in the linearized phase diagram \cite{Hallstedt2007}. The diffusion flux $\vec{J}$ follows Fick's equation as
\begin{equation}
\vec{J} = -D \nabla c
\label{Eq_J}
\end{equation}
where $D$ is the diffusion coefficient. The chemical driving force $\Delta g_{\alpha\beta}^\mathrm{CH}$ results in
\begin{equation}
\label{Eq.FChem}
\Delta g_{\alpha\beta}^\mathrm{CH}=m\Delta S_0\left(c-c_\mathrm{eq}\right)
\end{equation}
with $m$ as the slope in the linearized phase diagram separating single matrix phase and two-phase region, $\Delta S_0$ as the entropy of formation from supersaturated Al-Li to $\delta'$ and $c_{eq}$ as the equilibrium concentration at a flat interface.

\subsection{Simulation procedure}
\label{Sec_Sim}

\begin{figure}
\begin{center}
\includegraphics[]{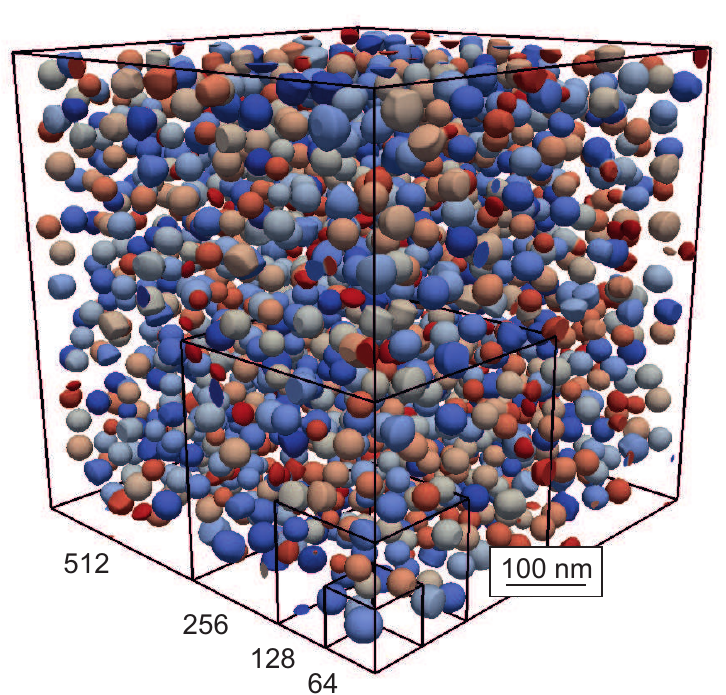}
\end{center}
\caption{A snapshot of reference microstructure (512$^3$ nm$^3$) with exemplary sample sizes (256$^3$ nm$^3$, 128$^3$ nm$^3$, 64$^3$ nm$^3$) is shown.}
\label{figsampling}
\end{figure}

A large-scale simulation with 512$^3$ grid points was conducted as the reference simulation box. This RVE was partitioned to samples of different size as shown in Figure \ref{figsampling}. An overview of the performed simulations can be seen in Table \ref{taboverview}. The samples were defined in different `classes' $v$ (in the following: \textit{64}, \textit{128}, \textit{256}, \textit{512} shown in left subscript) with corresponding cubic volumes $_vV$ (64$^3$ nm$^3$, 128$^3$ nm$^3$, 256$^3$ nm$^3$, 512$^3$ nm$^3$). The number of simulations for each class was defined as the quantity n$_s$ (see Tab. \ref{taboverview}). Periodic boundary conditions were applied. Precipitation, growth and ripening of $\delta '$ (stoichiometric Al$_3$Li) particles in an Al-9 at.\% Li alloy was considered. 

In total, 1924 precipitates were nucleated on random sites in the reference system in a stepwise manner during the first 250 $s$. This corresponds to a classical scenario were a high number of nucleation events occurs at the beginning, and then nucleation fades out. Nucleation sites and times were kept identical for all the samples. Elastic energy contributions due to the transformation were neglected for simplicity. Therefore, the precipitates had a generic spherical shape. Interface energy and interface mobility were 0.014~Jm$^{-2}$~\cite{Baumann1984} and 3$\times$10$^{-18}$ m$^{4}$J$^{-1}$s$^{-1}$, respectively. The diffusion coefficient of lithium in aluminium at the simulation temperature (473.15 K) was taken as 1.2$\times$10$^{-18}$ m$^2$s$^{-1}$~\cite{Callister2007}. The entropy of formation was $\Delta S_0=-9.7315 \times 10^5$ JK$^{-1}$m$^{-3}$ \cite{Chen1989} while the equilibrium Li concentration $c_{eq}$ was 6.67 at.\%. Time step and grid spacing were chosen as 0.25~s and 1~nm, respectively. All simulations used the same input parameters. The simulations were performed using a massively parallel version (Sec. \ref{Sec_Paral}) of the open source software OpenPhase \cite{OpenPhase,Tegeler2015,Tegeler2017}. The output of the simulation contains the evolution of individual precipitate volumes and densities. The calculations were performed on the clusters of the \textit{ICAMS} \cite{clustervulcan} and \textit{RWTH Aachen University} \cite{clusteraachen}. 

\subsection{MPI/OpenMP parallelization}
\label{Sec_Paral}
\begin{figure}
\begin{center}
\includegraphics[]{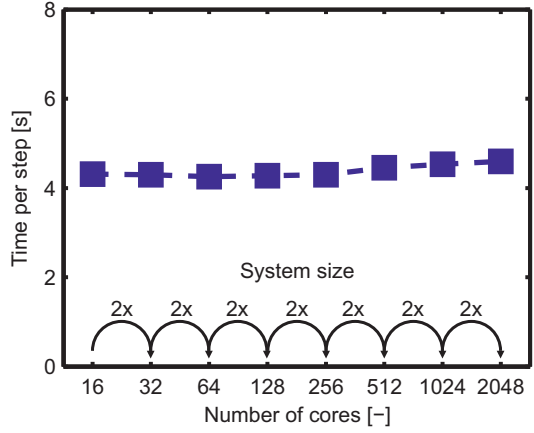}
\end{center}
  \caption{The performance-time-per-step is shown (blue squares) when the system size and the number of computation cores are doubled using parallelizing simulation techniques.}
\label{figscaling}
\end{figure}

The simulations have been performed in OpenPhase \cite{OpenPhase} using a hybrid-parallelization presented in \cite{Tegeler2015} and \cite{Tegeler2017}. To achieve hybrid parallelization, the message passage interface (MPI) standard and the open-multi-processing (OpenMP) application programming interface (API) were utilized. Distributed-parallelism provides the capabilities for large-scale simulations that overcome the restrictions of single node computers. This has been necessary to compute solutions for higher system sizes as their memory requirements exceeded the capacities of a single computational node. The parallelization uses a domain-decomposition with a wide halo approach \cite{Kjolstad2010} in order to avoid synchronization within a time step. A halo consists of a number of ghost cells stemming from neighbor domains that contain the information of the real cells. A halo allows more stencil-operations without communication. In the present case, nine additional layers were needed. Six layers are needed to average the driving force along the interface normal in order to recover the travelling wave solution necessary for the phase-field Equation~\ref{Eq.PFF}. One layer for marking grid points in or near the interface between phase fields. One layer for diffusion calculations and one layer for the computation of anti-trapping currents. The work on each sub-domain therefore increases as operations are duplicated on different processes. Hybrid-parallelization combining MPI and OpenMP is beneficial in this case as the number of processes is decreased, which in turn increases the size of sub-domains and reduces the ratio of ghost cells to inner cells. Figure~\ref{figscaling} shows the weak scaling performance of OpenPhase on the cluster in \textit{J{\"u}lich Research Centre, JuRoPA}, for a test using a uniform distribution of grains. Four blocks consisting of $50^3$ grid points were assigned to each MPI-process. The number of grains in the domain was $4 N_{\text{MPI}}$. As we see, the computation time remained the same when doubling both the number of cores and the system size.

\subsection{Methods of analysis}
\label{Sec_Ave}

In the simulations, the individual and average precipitate radius by assuming an equivalent sphere $_vR_j$=$\left(\frac{3}{4\pi}_vV_j\right)^{1/3}$ was tracked as
\begin{equation}
_v\langle R \rangle _i = \frac{1}{N}\sum_{j=1}^N \ _vR_j
\end{equation}

where $j$ runs over the precipitates inside a system $i$. The average over $m$ samples of the same class is calculated by $_v\langle R \rangle=\frac{1}{m}\sum_{i=1}^m$$_v\langle R \rangle _i$. Furthermore, the average precipitate centre-to-centre distance of every individual simulation $_v\langle d \rangle_i$ and of all sampling classes ($_v\langle d \rangle=\frac{1}{m} \sum_{i=1}^m$$\ _v\langle d \rangle_i$) can be measured and calculated using precipitate densities $_v\langle n \rangle _i$ and sample volumes $_vV$ assuming a homogeneous distribution as
\begin{equation}
_v\langle d \rangle _i = \sqrt[3]{\frac{_v\mathrm{V}}{_vn_i}} \quad.
\end{equation}

Precipitate radii and spacings are dimensionless when normalized by $\sqrt{Dt}$ (diffusion length) for generalization. The spacing characterizes the appropriate volume around each precipitate and therefore the precipitate number density.\\

In order to compare the results of sampled systems against the reference system and to give a measure of the accuracy, we computed the deviation $_v\delta_i$ (in $\%$) averaged over time for each individual sample $i$ of class $v$ as
\begin{equation}
\label{deltadev}
 _v\delta_i=\frac{1}{t_N}\sum_{t=1}^{t_N} \frac{\langle R \rangle_{t,i}-  {}_{ref}\langle R \rangle_{t}}{_{ref}\langle R \rangle_{t}} \times 100 \%
\end{equation}

where $t_N$ is the number of written outputs, $\langle R \rangle_{t,i}$ is the average precipitate radius at output time $t$ of sampling system $i$, and $ {}_{ref}\langle R \rangle_{t}$ is the average precipitate radius of the reference system at the same time. Thus, the average deviation $_v\Delta_{n_d}^{n_s}$ for the sub-set of combinations $n_d$ out of the set $n_s$ for specific sampling classes can be calculated as:
\begin{equation}
\label{Deltadev}
 _v\Delta_{n_d}^{n_s}=\frac{n_d!(n_s-n_d)!}{n_s!}\sum_{q_1=1}^{(n_s-(n_d-1))}\sum_{q_2=q_1+1}^{(n_s-(n_d-2))}...\sum_{q=n_d}^{(n_s-(n_d-n_d))}|\frac{1}{n_d}({}_v\delta_{q_1}+{}_v\delta_{q_2}+...+{}_v\delta_q)| \quad .
\end{equation}

The average precipitate spacing is compared to the reference system for individual classes $_v\omega_i$ and combinations $_v\Omega_{n_d}^{n_s}$ in the same way:
\begin{equation}
\label{omegadev}
 _v\omega_i=\frac{1}{t_N}\sum_{t=1}^{t_N} \frac{\langle d \rangle_{t,i} - {}_{ref}\langle d \rangle_{t}}{{}_{ref}\langle d \rangle_{t}} \times 100 \% \quad ,
\end{equation}
\begin{equation}
\label{Omegadec}
 _v\Omega_{n_d}^{n_s}=\frac{n_d!(n_s-n_d)!}{n_s!}\sum_{q_1=1}^{(n_s-(n_d-1))}\sum_{q_2=q_1+1}^{(n_s-(n_d-2))}...\sum_{q=n_d}^{(n_s-(n_d-n_d))}|\frac{1}{n_d}({}_v\omega_{q_1}+{}_v\omega_{q_2}+...+{}_v\omega_q)| \quad .
\end{equation}

The factorial term increases significantly with increasing $n_d$ and $n_s$, so only a few values can be calculated for a high number of samples for computational reasons. The computational effort needed for different simulation classes is assumed to scale linearly (Sec. \ref{Sec_Paral}), i.e. doubling the simulation size and computation power result in a constant simulation time. The variable $\lambda_{\mathrm{ref}}$ describes the computational effort (in \%) needed for a sample compared to the reference simulation ($\lambda_{\mathrm{ref}} = \frac{\mathrm{Total \ simulation \ time \ for \ a \ sample \ class}}{\mathrm{Total \ simulation \ time \ for \ the \ reference \ system}}$).\\

An investigation of the reliability of samples of different quantities $n_d$ requires the presentation of the deviation of precipitate sizes and spacings for each single combination $i$ of $n_d$ out of $n_s$ which are named by $\left(_va_{n_d}^{n_s}\right)_i$ and $\left(_vd_{n_d}^{n_s}\right)_i$, respectively. The mean size deviation for combinations having the same spacing deviation $_vd_{n_d}^{n_s}$ can be calculated by
\begin{equation}
\label{meana}
 _vm_{n_d}^{n_s}=\frac{1}{n_e}\sum_{q=1}^{n_e}{} \left(_va_{n_d}^{n_s}\right)_q
\end{equation}

where $n_e$ is the number of existing combinations for each spacing deviation $_vd_{n_d}^{n_s}$. The standard deviation $_vs_{n_d}^{n_s}$ for these mean values is given by
\begin{equation}
\label{stda}
 _vs_{n_d}^{n_s}=\sqrt[]{\frac{1}{n_e-1} \sum_{q=1}^{n_e}\left(\left(_va_{n_d}^{n_s}\right)_{q}-{}_vm_{n_d}^{n_s}\right)^2} \quad .
\end{equation}

\begin{table}
\begin{center}
  \caption{Overview of performed simulations with classes $v$, volumes $_vV$ (following Figure~\ref{figsampling}), quantities $n_s$ and reference covering information.}
\begin{tabular}{ c  c  c  c }
\hline
\label{taboverview}
  Class $v$ & Volume $_vV$ & Quantity $n_s$ & Reference covering [\%] \\
  \hline
  \hline
  512 & 512$^3$ nm$^3$ & 1 & -- \\
  256 & 256$^3$ nm$^3$ & 8 & 100\\
  128 & 128$^3$ nm$^3$ & 64 & 100\\
  64 & 64$^3$ nm$^3$  & 128 & 25\\
  \hline
\end{tabular}
\end{center}
\end{table}

\section{Simulation results}
\label{Sec_Res}

\begin{figure}
\begin{center}
\includegraphics[]{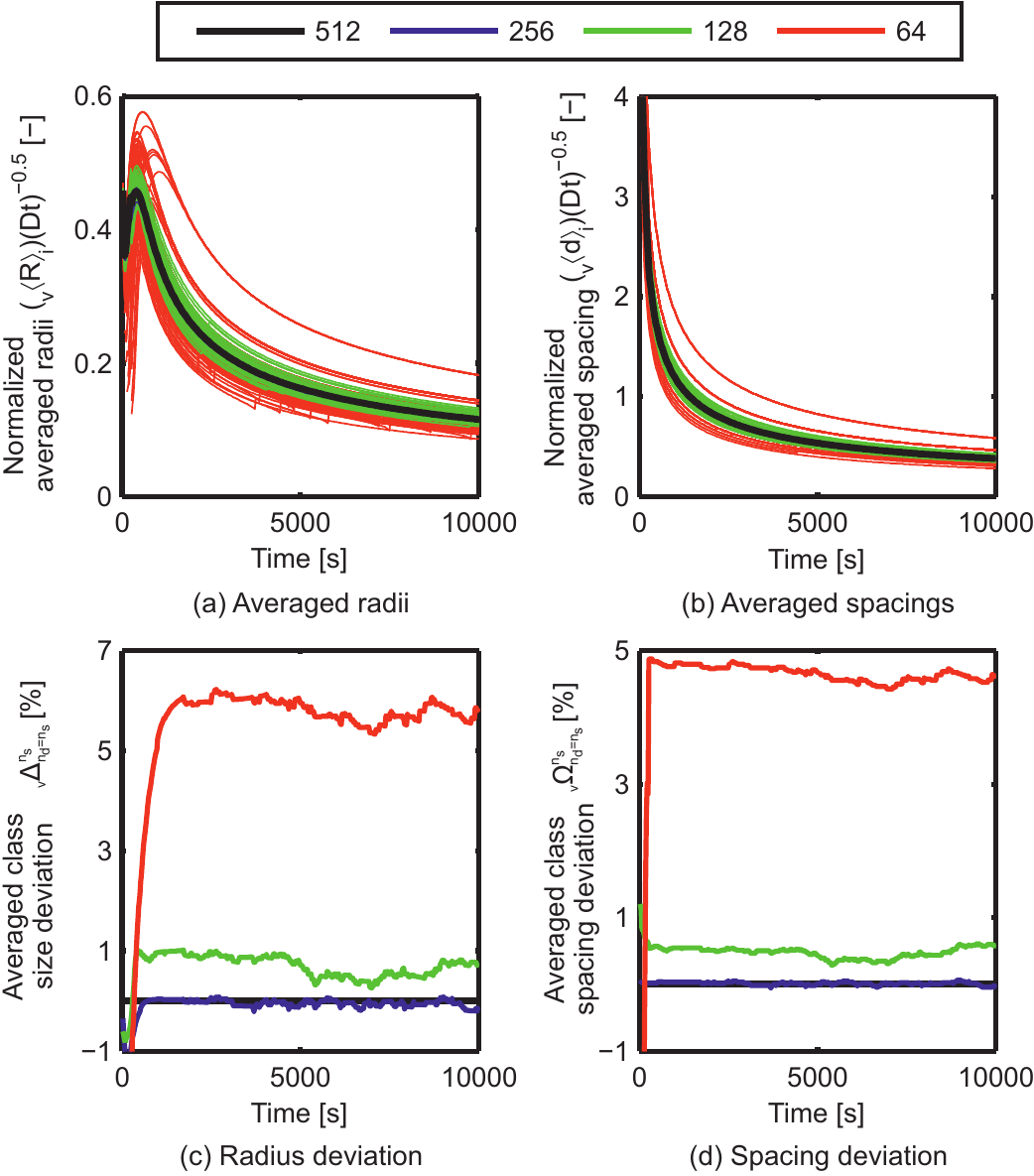}
\end{center}
\caption{Normalized (by diffusion length $\sqrt[]{Dt}$) averaged equivalent precipitate radii ${}_v\langle R \rangle_i$ for individual sampling systems $i$ (a) and deviation of the averaged sample classes $v$ with respect to the reference (c) as well as corrected averaged precipitate spacing ${}_v\langle d \rangle_i$ for individual sampling systems $i$ (b) and it's deviation from the reference for averaged sample classes $v$ with respect to the reference (d) are shown.}
\label{figprecsize}
\end{figure}

\begin{figure}
\begin{center}
\includegraphics[]{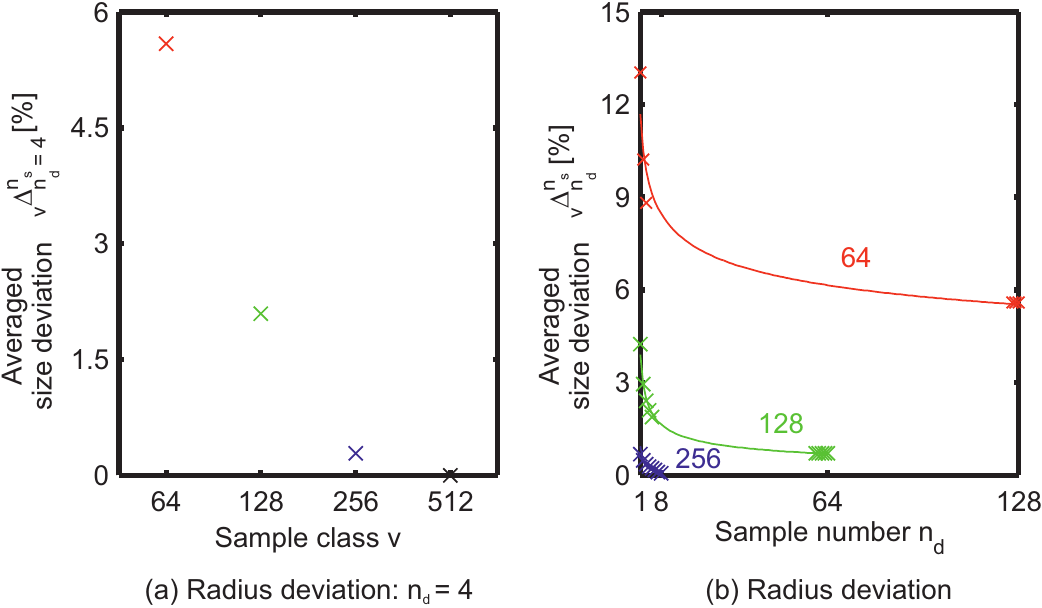}
\end{center}
\caption{Averaged size deviation $_v\Delta_{4}^{n_s}$ from the reference simulation for four samples out of the total number of available samples $n_s$ (a) and size deviation $_v\Delta_{n_d}^{n_s}$ from the reference simulation for different number of samples $n_d$ out of $n_s$ with fitted power laws (b) are shown. The fittings follow $y=ax^b$ with a=3.91 and b=-0.42 for \textit{128} and a=11.68 and b=-0.15 for \textit{64}.}
\label{figtime}
\end{figure}

Under the thermodynamic chemical driving force, the lithium-rich $\delta'$ precipitates started growing instantly within the supersaturated aluminium-lithium matrix. The size of the precipitates increased monotonically in the early stages of precipitation. Once the solute content in the matrix was depleted, precipitates evolve in a competitive ripening process during which larger precipitates grew at the expense of smaller ones. The precipitate size and spacing during the course of growth and ripening for each individual sample as well as the averaged size and spacing deviation of the full set of samples of each class are shown in Figures \ref{figprecsize}a-\ref{figprecsize}b and \ref{figprecsize}c-\ref{figprecsize}d, respectively. Both results show similar trends with respect to the system size. In general, it was found that for smaller simulation boxes the deviations with respect to the reference system are larger. The averaged size deviations as a function of system size (sample class) are compared in Figure \ref{figtime}a as well. The averaged deviations $_v\Delta_{n_d}^{n_s}$ and $_v\Omega_{n_d}^{n_s}$ of the sampling classes (Figure \ref{figprecsize}c and \ref{figprecsize}d) show that the results of the reference system are well recovered by sampling class \textit{256}. The sampling classes \textit{128} and \textit{64} show, however, larger deviations of about 1 and 6\%, respectively.\\

The deviation from the reference system was also found to be strongly dependent on the number of statistically averaged samples (Figure \ref{figtime}b). For each sample class, the deviation decays with a power-law when the number of statistically averaged samples increased. For the same coverage of the reference volume, it was found that smaller sampling classes showed larger deviations even for very large number of statistically averaged samples. For instance, a larger deviation is observed for sampling class \textit{128} at any number of statistically averaged samples although both 8 $\times$ \textit{256} and 64 $\times$ \textit{128} fully cover the reference system volume. This is attributed to the long-range nature of diffusion in the precipitation process which enables interaction with second- and higher-rank neighbouring precipitates.\\

\begin{figure}
\begin{center}
\includegraphics[]{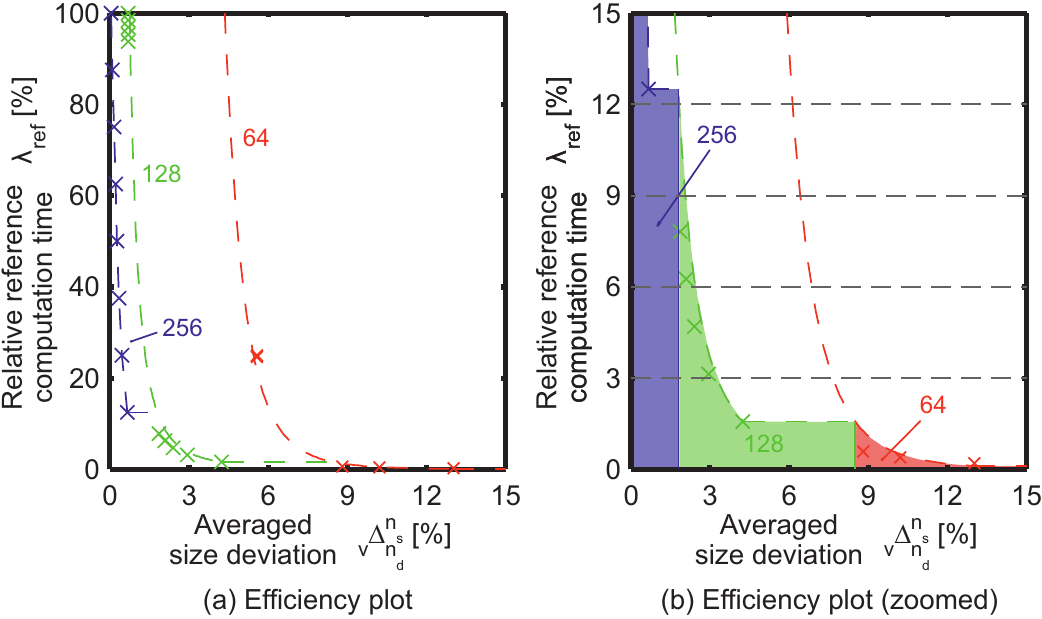}
\end{center}
\caption{Computational effort $\lambda_{\mathrm{ref}}$ (relative to the reference simulation) is mapped over averaged size deviation $_v\Delta_{n_d}^{n_s}$. Figure (b) shows zoomed in results of (a). The power law fittings follow $y=ax^b$ with a=52.35 and b=-2.4 for \textit{128} and a=987482.96 and b=-6.24 for \textit{64}.}
\label{simdevlinear}
\end{figure}

In order to study the trade-off between the computational costs and accuracy of the simulations, the total simulation time (computational effort) versus the deviations in different sampling classes was mapped in Figures \ref{simdevlinear} and the results are described in the following. The curves in Figure \ref{simdevlinear} are fitted to the function indicated in the caption. As it is shown in the map, for a linear scaling (Sec. \ref{Sec_Paral}) the smallest sample class (\textit{64}) showed to be only beneficial for deviations higher than 8.51 \%, whereas the bigger sampling class (\textit{128}) proved to be most efficient for deviations between 1.80 \% and 8.51 \% (Figure \ref{simdevlinear}). The biggest sampling class (\textit{256}) was only profitable in comparison to the other systems for deviations below 1.80 \%. It is evident that such a systematic mapping as presented here can be useful to decide the set-up of the sampling studies.

\section{Discussion}
\label{Sec_Discussion}
\subsection{Simulation evaluation}

In order to understand the sources of deviations in the results, one must pay particular attention to the effect of partitioning. In the current simulations, by to isolating a certain number of precipitates in a given volume from the rest of reference volume. This isolation of the precipitates leads to a variation in the local number density of the precipitates in individual simulation boxes. Since the simulations are conducted from the early stages of nucleation, the number density of the precipitates represents the average volume of supersaturated matrix per precipitate that determines the driving force for growth and ripening. Evidently, this effect is much more pronounced for smaller simulation boxes. For instance, in some of the simulation boxes in class \textit{64}, there are only one or two precipitates that can grow without competition from any other precipitate in the immediate neighbourhood. This results in a faster growth and therefore in larger deviations compared to the kinetics observed in the reference system. In fact, the red peaks in Figure \ref{figprecsize}a that associate with \textit{64} simulations indicate that larger precipitate radii are achievable in the smaller simulation boxes because of extra solute content (matrix) per precipitates number in some samples. This is a direct consequence of these precipitates having access to larger solute content in the given simulation box. \\

In the ripening stage, the effect of the local number density becomes much more complex because ripening is a competitive process that depends (in first-order approximation) on the relative size of the precipitates and thus, it is sensitive to the size distribution and spacing within each simulation box. During the course of simulations in the current study, the results showed that the number density of the precipitates in each simulation box provided information for anticipating the overall deviation in the results for both the growth and the ripening stage. This information can be used to perform more efficient sampling as will be discussed in Section \ref{Sec_RemIS}. It is important to note that in the later stages of ripening, the number density of precipitates may not provide enough information in the chosen RVE, as a wider neighbourhood around the precipitate may play a significant role in the ripening process. A controlled exchange of matter between samples by dynamic boundary conditions can overcome the bottleneck of specific efficient sample sizes at different stages of microstructure evolution.\\

\begin{table}[ht]
\caption{Calculated and measured fraction of precipitates sitting at simulation box boundaries facing a new environment after sampling due to periodic boundary conditions for different sampling classes $v$. }
\begin{center}
\begin{tabular}{ c  c  c }
\hline
\label{Tab.Fractions}
  Class $v$ & Calculated fraction [\%] & Measured fraction [\%] \\
  \hline
    \hline
  256 & 68.89 & 68.76 \\
  128 & 95.52 & 95.48 \\
  64  & 100 & 100 \\  
  \hline
\end{tabular}
\end{center}
\end{table}

Because of periodic boundary conditions in all simulations, the precipitates sitting close to the boundaries in sampled boxes experienced new interaction environment (neighbourhood) compared to their initial environment in the reference set-up. This can also be another source of deviation in the results. Table \ref{Tab.Fractions} presents the fraction of precipitates that interact with 'new' first-rank neighbours (compared to the reference system) owing to the periodic boundaries. The calculated and the measured values are listed. While all precipitates in class \textit{64} experienced new neighbourhoods, about 70\% of the precipitates in the largest class of sampling (\textit{256}) were located at the boundaries. This indicates that the new environment due to the periodic boundary conditions has a small effect on the deviation in the results. This is expected because of the random distribution of the precipitates in the reference system. Furthermore, the spacing between the precipitates that is evaluated in our simulations takes this effect also into account and provides a more accurate condition for choosing appropriate RVEs compared to the values for number density of the precipitates. Based on these results, we applied a concept of intelligent sampling such that effective number densities of sampled simulations matches closely with the reference experiment of interest. This concept is discussed in Section \ref{Sec_RemIS}.\\

\subsection{Sampling study using statistical analysis}
\label{Sec_Statistics}
The central observation from the simulation results is that the smaller the partitioned simulation box, the less accurate the statistical compilation of the results. These deviations originate from three specific size effects. The first effect is caused by the sampling of finite populations, for which the variance of a predicted average quantity, such as $\langle R_i \rangle$ or $\langle n_i \rangle$ generally increases with decreasing size of the individual samples (class size) that are statistically compiled. The second effect results from the confinement of the precipitates into isolated local groups interacting in a periodic simulation box without being able to exchange diffusive fluxes with far-field reservoirs. Consequently, the simulated coarsening rates could be altered. A third effect originates from the periodic boundary conditions which induce spatial correlations that affect the solute concentration field ahead of the (periodic images) of the precipitates. The simulation results proved that decomposing the microstructure into smaller disjoint sets did not considerably impact the accuracy of the simulations up to a certain limit. Space decomposition has been analysed in the past \cite{Kanit2003,Qidwai_2012} in particular for crystal plasticity and finite element formulations, where this kind of simplified microstructure is deemed as a weighted set of statistical volume elements (WSVEs). The rules for the selection of these WSVEs were established by Qidwai et al. \cite{Qidwai_2012} by utilizing two-point statistics \cite{Niezgoda2011} in CP-FEM simulations of plane strain deformation. The selection or definition of the solitary units in the case of precipitation is more complicated because diffusion may impose a long-range and long-lasting heterogeneity on the microstructure specially in the case, when grain boundary and pipe diffusion are considered. \\

Figure \ref{figtime} substantiates that within a certain accuracy it is possible to simulate precipitation by utilizing solitary units as a WSVE. The statistical analysis of the set-up of the simulations offered some insights into the conditions necessary to define the minimal/optimal size of the solitary units. It is stressed that such analysis must be performed \emph{a priori} to avoid wasting computational resources during the simulations. \\

\begin{figure}[ht]
\begin{center}
\includegraphics[]{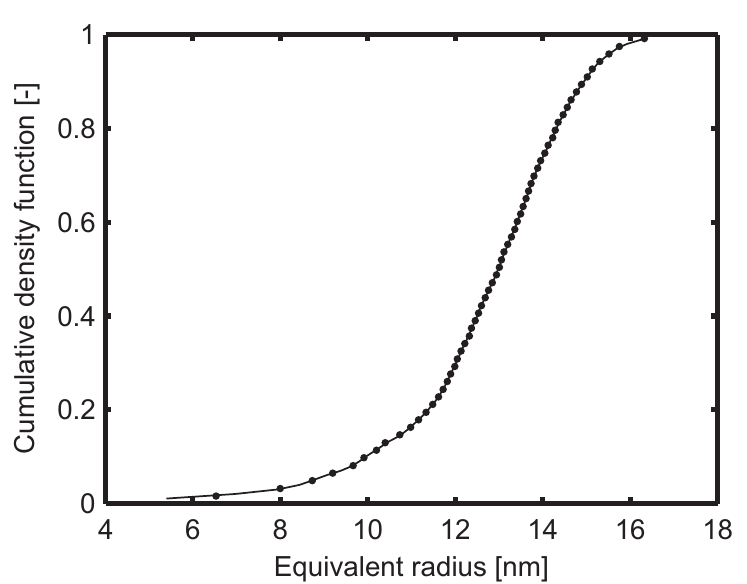}
\end{center}
\caption{The empirical cumulative distribution (solid black line) and the kernel-density estimate (black dots) are shown for the last time step (40000) in the reference simulation box. The kernel width is equal to 0.1. Comparing the empirical cumulative distribution with its kernel-density estimate shows the applicability of sampling the replica ensemble with the kernel density estimated CDF.}
\label{fitLastTimestep}
\end{figure}

To begin with, the final population in the reference simulation box was analysed. The box contained after the final simulated step (40000 time steps) 1845 particles. In Figure \ref{fitLastTimestep} the calculated cumulative density function (CDF) was fitted using a kernel density estimator from Matlab\textsuperscript{\textregistered} with a Gaussian kernel (kernel width=0.1). \\

\begin{figure}[ht]
\begin{center}
\includegraphics[]{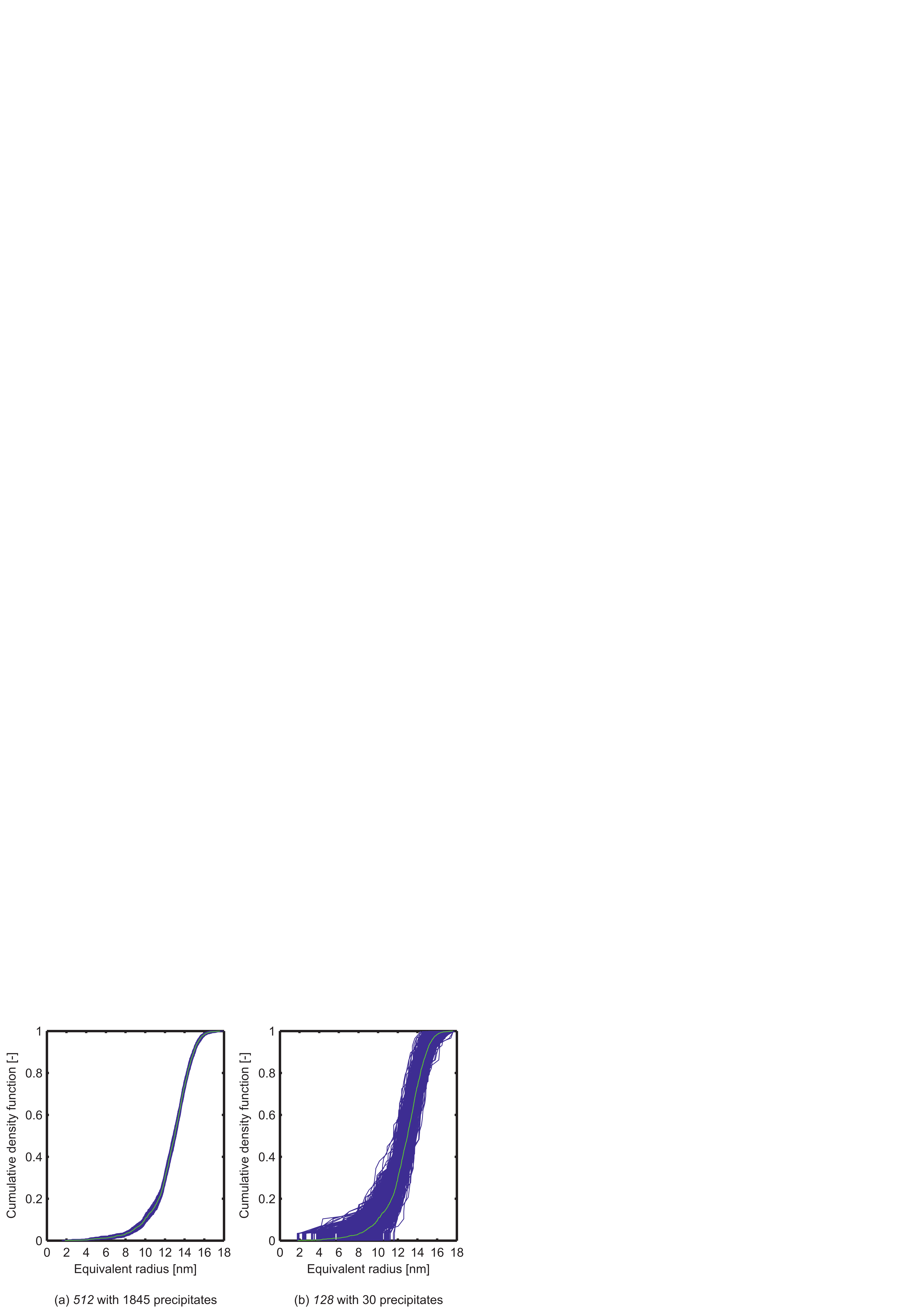}
\end{center}
\caption{Replica sampling for different system sizes are plotted. (a) For 512$^3$~nm$^3$ and 1845 precipitates (reference system) sampled based on the fitted CDF from Figure \ref{fitLastTimestep} the CDF is very similar for all replicas. (b) For a sample with size 128$^3$~nm$^3$ and 30 particles the scatter in the CDF curves substantially increases.}
\label{Subsystems}
\end{figure}

In a first analysis, 1000 replicas containing 1845 precipitate each were utilized to represent the microstructure. The spatial distribution of the precipitates  were random. The size of the precipitates were defined by sampling the inverse CDF, Figure \ref{fitLastTimestep}. In Figure \ref{Subsystems}a, realizations of the possible CDFs are plotted. The variation of the individual CDF depicted as the width of the deviations from the solid line was found to be small. In a sample of class \textit{128} with equal particle density only 30 particles are hosted on average. We followed the same sampling procedure to analyse the corresponding variation in the cumulative distribution (Figure \ref{Subsystems}b) in the smaller system. The expected CDF is independent of the system size, whereas the variance depends strongly on the number of sampled particles. \\

\begin{figure}[ht]
\begin{center}
\includegraphics[]{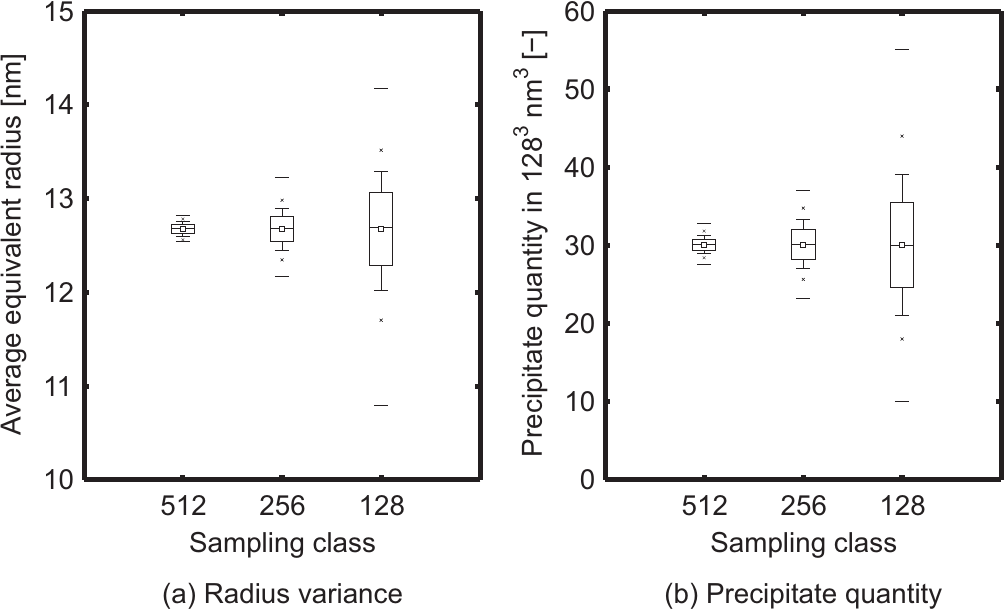}
\end{center}
\caption{(a) Mean particle radius in the replica sample is shown. The variation of the mean increases with reducing the replica size, whereas the expected value is constant. (b) The deviation of the replicas depending on the replica size (mean, standard deviation, variance) is shown.}
\label{fig8}
\end{figure}

To better understand the causes of the variation, the mean particle size compiled for each replica study is plotted in Figure \ref{fig8}a. Each dot represents the average particle size in the system. For a system with a size of $512^3$ nm$^3$ the average particles size has a very low variance, which is a direct consequence of Figure \ref{Subsystems}a. By decreasing the system size, the variance of the mean increased as less particles were hosted. By simulating more and more replicas, an individual realization far away from the mean is more likely. However, the expected mean radius is constant and independent of the system size (Figure~\ref{fig8}b). For these reasons, the differences in radius and number density in the phase-field simulations are certainly a consequence of the finite population sampled in smaller systems. It is stressed that this last conclusion is only valid for distributions given in a random or close to random spatial arrangement. The conditions for an efficient partitioning of a non-random distribution will evidently differ. Nevertheless, the significance of this finding is the proof that the same practical prediction accuracy can be obtained by splitting the reference phase-field simulation box into an ensemble of individual but much smaller and foremost independent simulation samples. This opens the possibility to their significantly faster solution by improving the accessing of memory in the algorithms as has been done, for example, for the simulation of primary recrystallization \cite{Kuehbach2016}. In the case of more complex particle distributions or in cases where diffusion plays a more prominent role, the use of methods inspired in data-science can be of use \cite{Steinmetz2016}. For instance, two-point statistics with consideration of the concentration profiles can be used.\\

An inconvenience of the partitioning is that an optimal sub-division strategy must be designed before the simulations are even performed to avoid wasting resources. The realization of precipitate number density as primary controlling parameter in these phase-field simulations devises a tool for optimizing the simulations. In the next section, a concept of intelligent sampling is introduced as a possible strategy to deal with this problem.

\subsection{Intelligent sampling}
\label{Sec_RemIS}
In order to achieve a better trade-off ratio between accuracy and computational costs of the sampling, it is logical to replace the random sampling process with a computationally-guided set of samples that represents the reference RVE more accurately. The findings of the current study indicate that the precipitate spacing is the primary controlling parameter which influences the kinetics of growth and ripening in our simulations. In this section, we compare the results of random sampling with samples which are intelligently chosen by considering their precipitate spacing index. \\

\begin{figure}
\begin{center}
\includegraphics[]{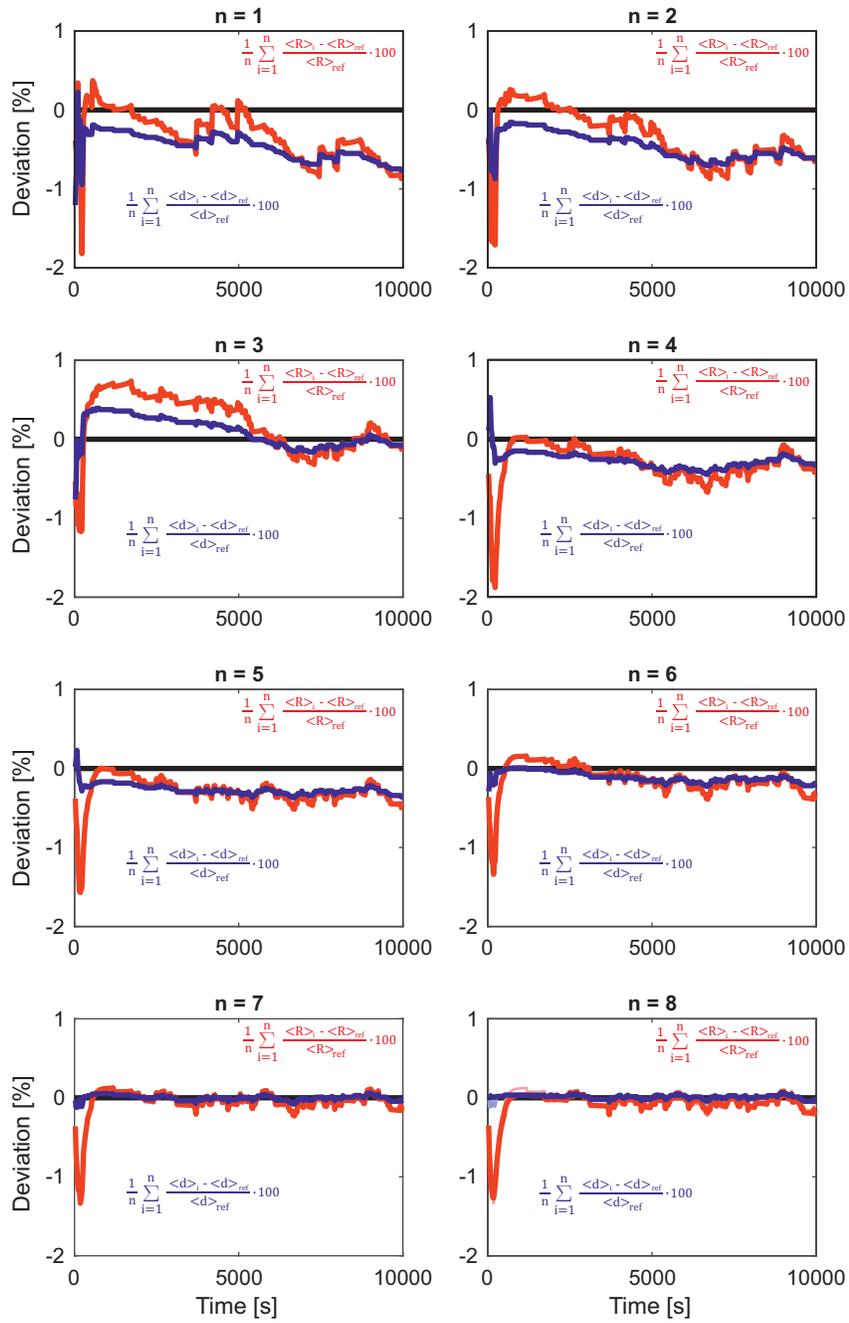}
\end{center}
\caption{Precipitate size (red) and spacing (blue) deviation evolution of class \textit{256} from the reference simulation for different sample quantities $n$ (following Tab. \ref{Tab.deviation}) are plotted.}
\label{Fig.Li-A6}
\end{figure}

\begin{figure}
\begin{center}
\includegraphics[]{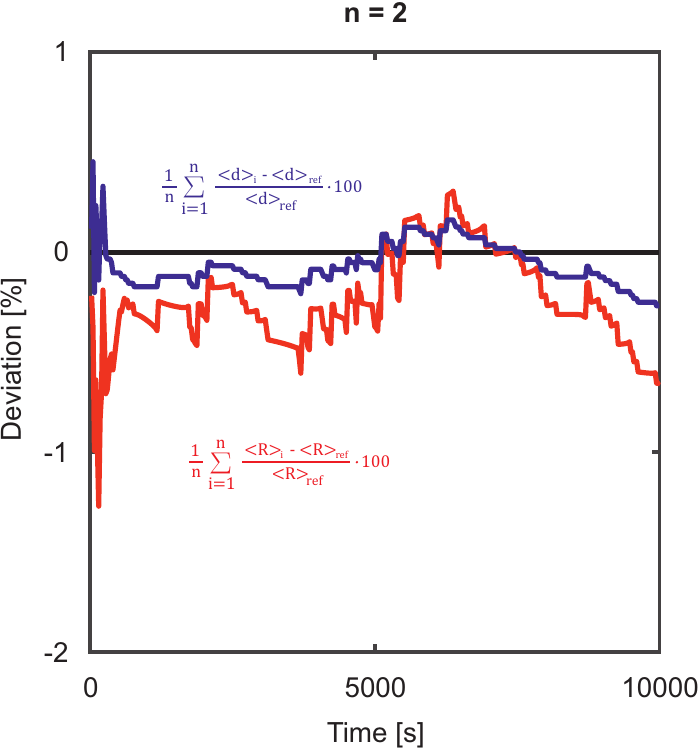}
\end{center}
\caption{Precipitate size (red) and spacing (blue) deviation evolution from the reference simulation for intelligent sampling using samples 4 and 8 from Table \ref{Tab.deviation} ($\frac{1}{2}\sum_{j=1}^{2} {}_{256}\omega_j^\mathrm{init}=-0.00005 \%$) are shown. The averaged radius deviation is lower (-0.23\%) than for random sampling (see $n=2$ in Figure~\ref{Fig.Li-A6}: -0.33\%).}
\label{Fig.Li-A7}
\end{figure}

Table \ref{Tab.deviation} lists the deviations in precipitate spacing (compared the reference system) after initialization (at 250~s) for all 8 samples in class \textit{256}. The samples were merged stepwise up to the full spatial covering ($n=8$) of the reference volume. By comparison (Figure~\ref{Fig.Li-A6}), it is found that a higher accuracy is achieved when the gap of the precipitate spacing is closed. Hence, choosing those samples which have a closer spacing index compared to the reference system represents closer number density of the precipitates and reduces the amount of final deviation. The deviation for an intelligently chosen ensemble of samples (Figure~\ref{Fig.Li-A7}) with n$_d$=2 and initial spacing deviation of -0.00005\% yields an average size deviation of -0.23\% which is lower than a random ensemble of the same size (initial spacing deviation: -0.28\%) as shown in Figure~\ref{Fig.Li-A6} and Table \ref{Tab.deviation} with in a deviation of -0.33\%.\\

\begin{figure}
\begin{center}
\includegraphics[]{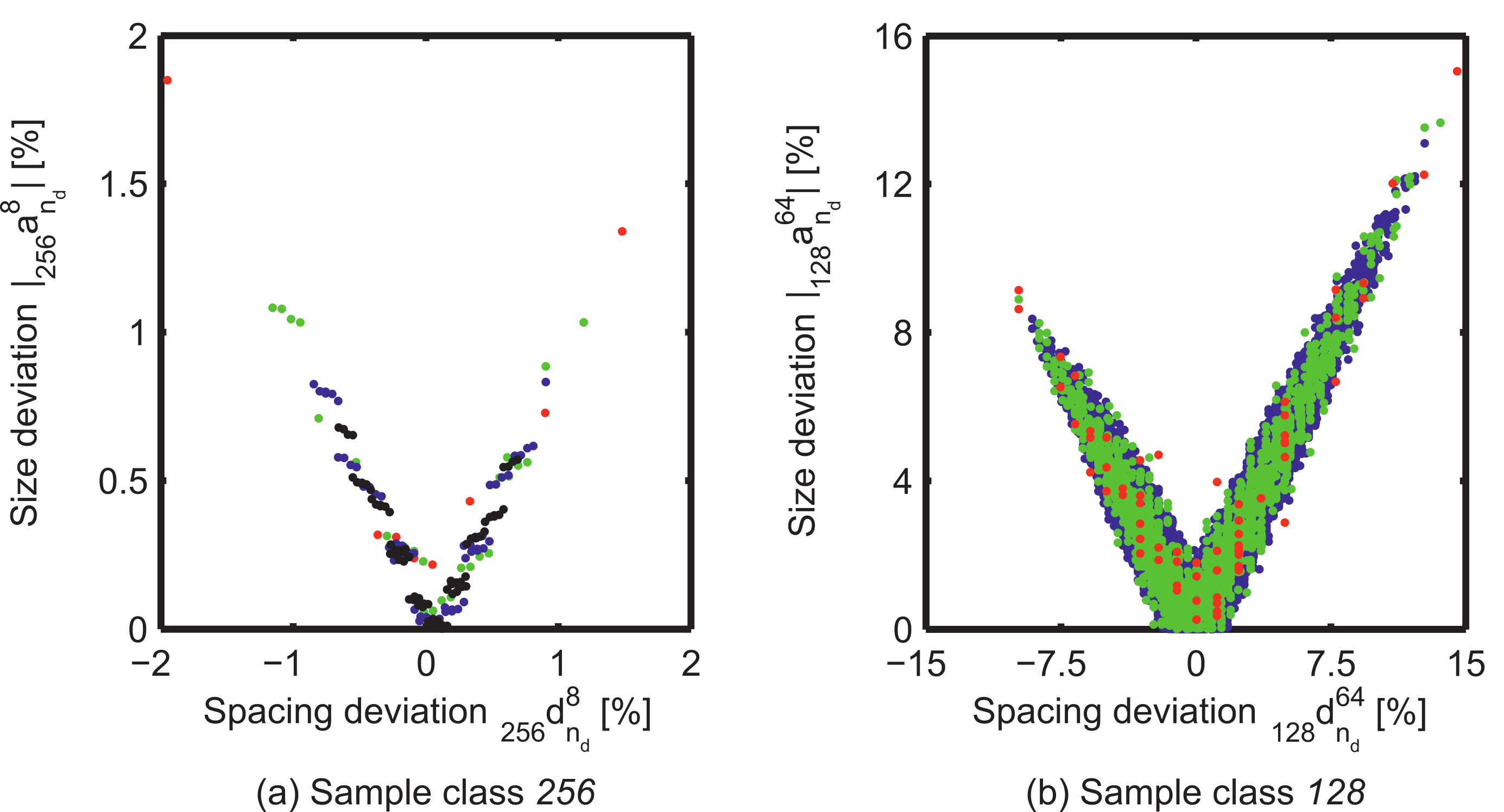}
\end{center}
\caption{Absolute size deviation $_{v}a_{n_d}^{n_s}$ plotted against spacing deviation $_{v}d_{n_d}^{n_s}$ for each individual available sample combination for \textit{256} (a) and \textit{128} (b). Colour scheme: $n_d$ = 1: red, $n_d$ = 2: green, $n_d$ = 3: blue, $n_d$ = 4: black.}
\label{intsam1}
\end{figure}

\begin{figure}
\begin{center}
\includegraphics[]{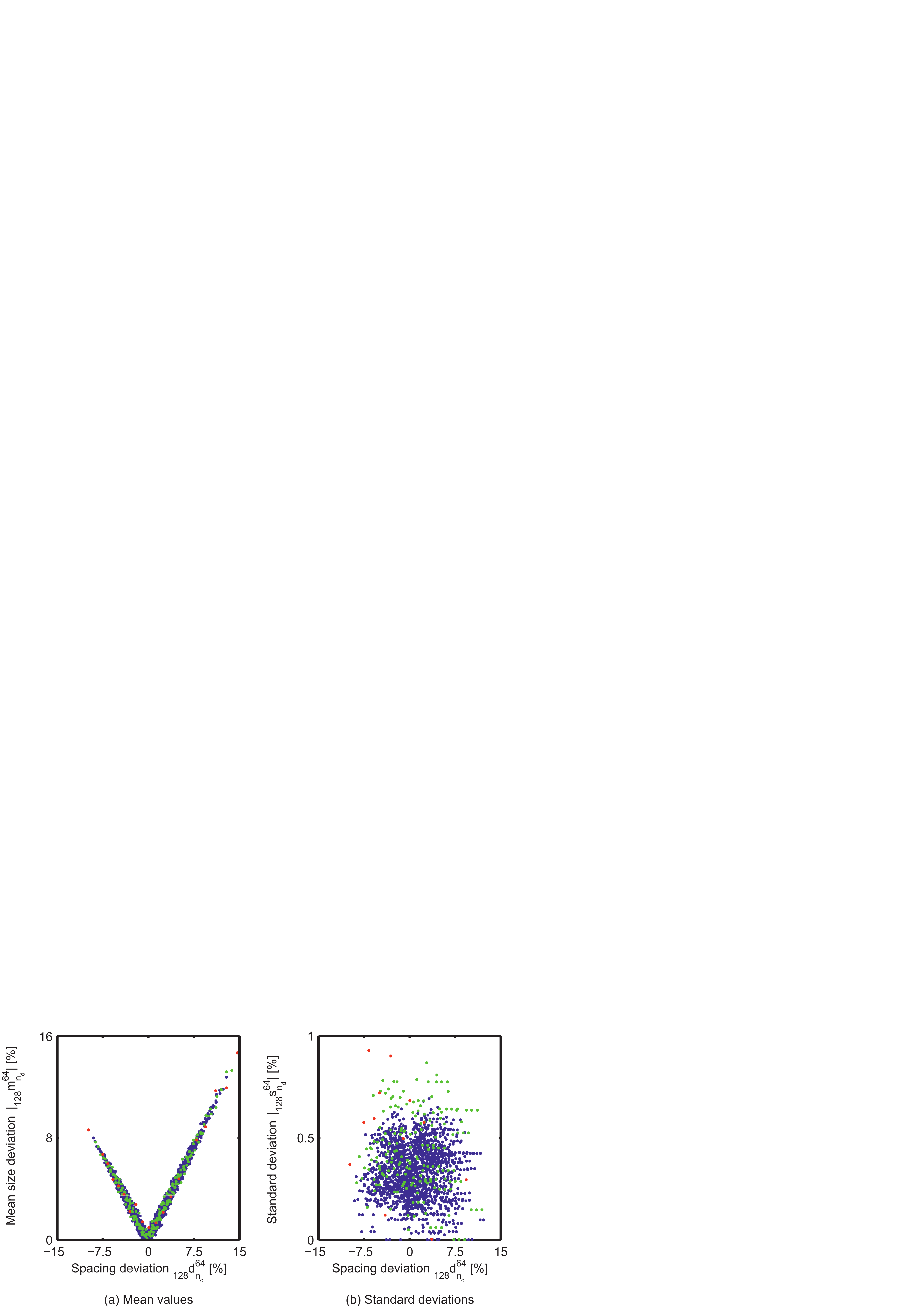}
\end{center}
\caption{Absolute mean size deviations $_{128}m_{n_d}^{64}$ (a) and absolute standard deviations $_{128}s_{n_d}^{64}$ (b) plotted against identical values of spacing deviations $_{128}d_{n_d}^{64}$ in \textit{128}. Colour scheme: $n_d$ = 1: red, $n_d$ = 2: green, $n_d$ = 3: blue.}
\label{intsam2}
\end{figure}

A systematic analysis was conducted by combining different numbers of samples ($n_d$ per combination $i$: 1, 2, 3, 4) and plotting their individual deviation in precipitate spacing $\left(_va_{n_d}^{8}\right)_i$ against their size deviation $\left(_vd_{n_d}^{8}\right)_i$. The results for classes \textit{256} and \textit{128} (Figure~\ref{intsam1}a and \ref{intsam1}b) show that the maximum and minimum values are reduced by increasing the sample number $n_d$. While the deviations revealed an almost symmetric behaviour for \textit{256} (Figure~\ref{intsam1}a), they behaved asymmetrically for \textit{128} (Figure~\ref{intsam1}b) by a shift in the balance point to the right. This is consistent with the deviations observed in Figure \ref{figprecsize}c. Figures \ref{intsam2}a and \ref{intsam2}b depict the mean and the standard deviation for sample class \textit{128} (sample class \textit{256} does not give enough data points for mean and standard deviation calculation). These findings evidence that sampling simulations with spacing properties comparable to the reference simulation give rise to only small size deviations. Thus, the total computation costs can be decreased by an intelligent choice of the samples. Naturally, the results and their reliability can be improved by increasing the number of samples $n_d$ in each ensemble (Figure~\ref{intsam2}a and \ref{intsam2}b). \\

\begin{table}[ht]
\caption{Spacing deviation $_{256}\omega_i^\mathrm{init}$ in the precipitate spacing with respect to the reference system after initialization sequence (250 s) for available samples of class \textit{256} and deviations after stepwise additive averaging of samples.}
\begin{center}
\begin{tabular}{ c ||  c  c  c  c  c  c  c  c }
\hline
\label{Tab.deviation}
  $i$ & 1 & 2 & 3 & 4 & 5 & 6 & 7 & 8\\
  $_{256}\omega_i^\mathrm{init}$ in [\%] & --0.21 & --0.34 & +1.50 & --0.07 & --1.93 & +0.77 & +0.35 & +0.07 \\
    $\frac{1}{n}\sum_{j=1}^{n=i} {}_{256}\omega_j^\mathrm{init}$ in [\%] & --0.21 & --0.28 & +0.32 & +0.22 & --0.21 & --0.05 & +0.01 & +0.02 \\
  \hline
\end{tabular}
\end{center}
\end{table}

\subsection{Remarks}

In this study, we have discussed the choice of an adequate RVE in a long-range diffusion-controlled process but with a rather homogeneous nature. While the results of the current investigations indicate one primary controlling parameter, i.e. precipitate spacing, one may expect a far more complex situation for a heterogeneous microstructure. In particular, systems which include multiphysics of different length and time scales may challenge the sampling and statistical averaging method proposed in this study. Nevertheless, the concept of replacing large RVEs by smaller independent simulation boxes remains valid. For each specific problem, the trade-off between computational costs and accuracy of the sampling can be mapped for future reference. This can be further enhanced by performing intelligent sampling which enables systematic reproduction of the reference system. This is possible by characterizing the primary controlling parameter(s) of the problem in hand. Instead of using a reference set-up, intelligent sampling can be performed by generating adequate samples and connecting these individual samples with each other by introducing dynamic boundaries to achieve reliable sampling accuracies across different stages (and interaction ranges) of microstructure evolution. An advantage of the current proposed method is the possibility for performing parallel independent studies of the sampled sub-systems with significantly reduced time-to-solution for any problem.\\

\section{Conclusions}
\label{Sec_Conc}
\begin{itemize}
\item Precipitation and growth of $\delta '$ precipitates in Al-9 at.\% Li alloy was studied by means of large-scale phase-field simulations. A large-scale reference simulation was incrementally partitioned into smaller samples and simulated independently.

\item The results of the simulations evinced that the partitioning of the reference simulation domain up to a well-defined limit negligibility impact the accuracy of the simulations compared to the reference simulation.

\item The cause of the deviation was traced back to sampling effects stemming from the neighbouring topology of precipitation. For the case studied, long-range diffusion did not seem to affect the accuracy. This may not be the case in long-time ripening or for a more complex precipitation topology as may be observed in case of particle clustering.

\item The number density of the precipitate has been characterized as the primary controlling parameter in the simulations. Based on this observation, a concept for intelligent sampling has been proposed that allows a computationally-efficient simulation procedure. The advantage of the proposed method is that it allows an a-priori definition of the optimal sub-system size with the inherent saving of computational resources.

\item  The more important conclusion of this study is that, with a careful selection of the sub-system size and partitioning, the method can be extended to more heterogeneous systems. The conditions of optimal partitioning will be the issue of future research.
\end{itemize}

\section*{Acknowledgements}
The authors gratefully acknowledge the financial support from the Deutsche Forschungsgemeinschaft (DFG) within the \emph{Reinhart Koselleck-Project} (GO 335/44-1). RDK acknowledges the financial support from DFG for his Eigene Stelle under the project DA 1655/1-1. The simulations were partly performed on the RWTH Aachen University computing cluster within the scope of the JARAHPC project JARA0076.

\bibliography{References}

\end{document}